\documentclass[12pt]{article}
\usepackage{amsmath,amssymb}
\newtheorem{lem}{Lemma}
\newtheorem{thm}{Theorem}

\newcommand{\beqn}{\begin{eqnarray}}
\newcommand{\eeqn}{\end{eqnarray}}
\newcommand{\rd}{\partial}
\newcommand{\ep}{\varepsilon}
\newcommand{\psv}{\overrightarrow{\psi}}
\newcommand{\C}{{\mathbf{C}}}
\newcommand{\Z}{{\mathbf{Z}}}

\renewcommand{\sl}{\mathfrak{sl}}

\newcommand{\g}{{\mathfrak{g}}}
\newcommand{\s}{{\mathfrak{s}}}
\newcommand{\h}{{\mathfrak{h}}}

\renewcommand{\b}{{\mathfrak{b}}}
\newcommand{\diag}{{\mathrm{diag}}}

\newcommand{\al}{\alpha}

\makeatletter
\newenvironment{proof}[1][\proofname]{\par
  \normalfont
  \topsep6\p@\@plus6\p@ \trivlist
\item[\hskip\labelsep{\bfseries #1}\@addpunct{.}]\ignorespaces
  }{%
  \endtrivlist
  }
\newcommand{\proofname}{Proof}
\makeatother

\makeatletter
\def\qedsymbol{\RIfM@\bgroup\else$\bgroup\aftergroup$\fi
  \vcenter{\hrule\hbox{\vrule\@height.8em\kern.6em\vrule}\hrule}\egroup}
\def\qed{\RIfM@\else\unskip\nobreak\fi\quad\qedsymbol}
\makeatother

\begin{document}
\title
{Similarity reduction of the modified \\ Yajima-Oikawa equation}
\author
{Tetsuya KIKUCHI${}^1$\thanks{{\it E-mail address} 
: tkikuchi@math.tohoku.ac.jp,},
Takeshi IKEDA${}^2$\thanks{{\it E-mail address} : ike@xmath.ous.ac.jp}
and
Saburo KAKEI${}^3$\thanks{{\it E-mail address} : kakei@rkmath.rikkyo.ac.jp}}
\date{August 29, 2003}

\maketitle
\centerline{${}^1${\it Mathematical Institute, Tohoku University,}}
\centerline{\it Sendai 980-8578, JAPAN}
\medskip
\centerline{${}^2$
{\it Department of Applied Mathematics, Okayama University of Science, }}
\centerline{\it Okayama 700-0005, JAPAN}
\medskip
\centerline{${}^3${\it Department of Mathematics, Rikkyo University,}}
\centerline{\it Tokyo 171-8501, JAPAN}

\begin{abstract}
  We study a similarity reduction of the modified Yajima-Oikawa hierarchy.
  The hierarchy is associated with a non-standard Heisenberg subalgebra
  in the affine Lie algebra of type $A_2^{(1)}.$
  The system of equations for self-similar solutions
  is presented as a Hamiltonian system of degree of freedom
  two, and admits a group of B\"acklund transformations
  isomorphic to the affine Weyl group of type $A_2^{(1)}$.
  We show that the system is equivalent to a two-parameter 
  family of the fifth Painlev\'e equation.
\end{abstract}

\section{Introduction}
\setcounter{equation}{0}
In applications of the theory of affine Lie algebras to integrable
hierarchies, the {\it Heisenberg subalgebras}  play important roles,
since they correspond to the varieties of time-evolutions.
Let $\hat{\g}$ be the untwisted affine Lie algebra associated with a
finite-dimensional
simple Lie algebra $\g.$
Up to conjugacy, the Heisenberg subalgebras in $\hat{\g}$ are in 
one-to-one correspondence
with the conjugacy classes of
the Weyl group of $\g$
\cite{112}.
In particular, the conjugacy class containing the Coxeter element, to which
the {\it principal\/} Heisenberg subalgebra of $\hat{\g}$ is
associated, leads to the Drinfel'd-Sokolov hierarchy \cite{DS}, whereas
the class of the identity element corresponds to the {\it
  homogeneous\/} Heisenberg subalgebra.
Associated with arbitrary conjugacy class,
M. F. de Groot, T. J. Hollowood, J. L. Miramontes \cite{gds1}
developed the theory of integrable systems
called generalized Drinfel'd-Sokolov hierarchies.

When $\g$ is of type $A_{n-1}$, the conjugacy
classes are parametrized by the partitions of $n.$
In this paper we consider the {\it modified Yajima-Oikawa hierarchy}, which
turns out to be a hierarchy related to
the affine Lie algebra of type $A_2^{(1)}$ and its non-standard Heisenberg
subalgebra associated with the partition $(2,1),$
while the {\it principal\/} (resp. {\it
homogeneous\/} ) case corresponds to the partition $(3)$ (resp. $(1,1,1)$ ).

Among the issues on integrable hierarchies,
the study of similarity reduction is important.
For example,
M. Noumi and Y. Yamada 
introduced a higher order Painlev\'e system associated with
the affine root system of type $A_{n-1}^{(1)}$ 
\cite{Noumi-Yamada:A} and
now the system
is known to be equivalent to a similarity reduction of the
system associated with the Coxeter 
class $(n)$ of $A_{n-1}$.
The aim of this paper is to investigate a similarity reduction of the
modified Yajima-Oikawa hierarchy.
Starting with universal viewpoints, 
we derive a system of ordinary differential equations
for unknown functions $f_0, f_1, f_2, u_0, u_1, u_2, g, q, r$ 
and complex paremeters $\alpha_0, \alpha_1, \alpha_2$:
\beqn
\begin{array}{ll}
  \alpha_0' = \alpha_1' = \alpha_2' = 0,   &   \\
    f_0' = f_0(u_2-u_0)-\alpha_0, & g'
    =g(u_0-u_2)-qf_1+rf_2 + \alpha_0 + 4,        \\
      f_1'=f_1(u_0-u_1)-r\alpha_1,    &
    3q'=3q(u_1-u_0)+qf_0-f_2,                    \\
      f_2'=f_2(u_1-u_2)-q\alpha_2,    &
      3r'=3r(u_2-u_1)-rf_0+f_1.
\label{symmetricform}
\end{array}
\eeqn
where $' = d/dx$ denote the derivative with respect to the
independent variable $x$.
Under the algebraic relations
\begin{equation}
\begin{array}{c}
 \alpha_0 + \alpha_1 + \alpha_2 = -4, \quad
 g = f_0 + 3qr, \quad u_0 + u_1 + u_2 = 0, \quad
 u_1 = qr, \\
2gu_0 = qf_1 - rf_2 - gqr - \alpha_0 -2,
\end{array}
\label{algebraicrelations}
\end{equation}
the system \eqref{symmetricform} 
turns out to be equivalent to
the fifth Painlev\'e equation for $y= -f_0/(3u_1)$:
\begin{align*}
   y'' =& \left(\frac{1}{2y}
          + \frac{1}{y-1} \right) (y')^2
          - \frac{y'}{x}
          + \frac{(y-1)^2}{x^2} \left(Ay + \frac{B}{y} \right)
          + \frac{C}{x}y
          + D\frac{y(y+1)}{(y-1)},
\end{align*}
where the change of variable $x \to x^2$ is employed and
the parameters are given by
$$
 A = \frac{1}{2}\left(\frac{\alpha_2 - \alpha_1}{12}\right)^2,
\quad
 B = -\frac{1}{2}\left(\frac{\alpha_0}{4}\right)^2,
\quad
 C = -\frac{\alpha_2-\alpha_1}{18},
\quad
 D = -\frac{1}{18}.
$$
On introducing the system \eqref{symmetricform},
we shall describe the system 
in three ways:

\medskip
1. Compatibility condition for a system of linear differential
equations (Section \ref{Laxpair}),

2. A Hamiltonian system whose degree of freedom is two 
(Theorem \ref{thm:Hamil}), 

3. Hirota bilinear equations for $\tau$-functions
(Theorem \ref{thm:bilinear}).

\medskip

The system \eqref{symmetricform} has a symmetry of
the affine Weyl group of type $A_2^{(1)}$ as a
group of B\"acklund transformations.
First we give the symmetry as the compatibility
of gauge transformations of linear differential
equations and state it in the automorphism of 
the differential field 
$$
K = \C(\alpha_0,\alpha_1,\alpha_2,f_0,f_1,f_2,g,q,r,u_0,u_1,u_2)
$$ 
with the derivation $':K \to K$ defined by \eqref{symmetricform}
and algebraic relations \eqref{algebraicrelations}
(Theorem \ref{thm:Weyl}).
Then we extend the action of affine Weyl group 
on $K$ to the extended field $\widehat{F}$ of $K$:
$$
\widehat{F} = \C(\alpha_0, \alpha_1, \alpha_2,x; \tau_0,\tau_1,\tau_2,
\sigma_1,\sigma_2,\tau_0',\tau_1',\tau_2',\sigma_1',\sigma_2')
$$
as a B\"acklund transformations,
which is discussed in section \ref{WeylTau} (Theorem \ref{last}).

The paper is organized as follows.
In Sect.\ref{Liealg}, we review the notation
related to the affine Lie algebra of type $A_2^{(1)}.$
On the basis of the affine Lie algebra, we introduce
the modified Yajima-Oikawa hierarchy in Sect.\ref{hier}.
In Sect.\ref{Similarity}, we consider a condition
of self-similarity on the solutions of the hierarchy.
This condition yields a system of ordinary diferential
equations, which is a main object in this paper.
In Sect.\ref{Laxpair}, the condition of
self-similarity is also presented as a Lax-type equation.
In Sect.\ref{Backlund}, we give a Weyl group symmetry
of the system as a gauge transformation
of the Lax equatoin (Theorem \ref{thm:Weyl}).
In Sect.\ref{Hamil} a Hamiltonian structure is introduced
(Theorem \ref{thm:Hamil}).
In Sect.\ref{Painleve} we prove
that our system
is equivalent to a two-parameter family
of the fifth Painlev\'e equation.
In Sect.\ref{taufunctions} we introduce a set of $\tau$-functions
and give a bilinear form of differential system
(Theorem \ref{thm:bilinear}).
Then in Sect.\ref{Jacobi} we lift the action of Weyl group
to the $\tau$-functions (Theorem \ref{thm:tauaut}) and
give a Jacobi-Trudi type formula (\ref{jacobi-trudi})
for the Weyl group orbit of the $\tau$-functions.
In Sect.\ref{WeylTau}, we prove that the Weyl group
action on the $\tau$-functions commute with
the derivation $' = d/dx$.

\section{Preliminaries on the affine Lie algebra of type $A_2^{(1)}$}
\label{Liealg}
\setcounter{equation}{0}
In this section,
we collect necessary notions about the affine Lie algebra of type
$A_2^{(1)}.$
We mainly follow the notation used in \cite{Kacbook},
to which one should refer for further details.

Let $\g=\sl_3.$
The affine Lie algebra $\hat{\g}$ is realized as a central extension
of the loop algebra $L\g={\sl}_3 \left(\C[z,z^{-1}] \right)$, together with
the derivation $d=z\rd_z$ 
$$
\hat{\g}
=
{\sl}_3 \left( \C[z,z^{-1}] \right)\oplus
\C c \oplus \C d,
$$
where $c$ denotes the canonical central element.
Let us define the Chevalley generators $E_i,F_i,H_i (i=0,1,2)$ for
the affine Lie algebra $\hat{\g}$ by
\begin{equation}
E_0 = z E_{3,1},\;
E_1 =   E_{1,2},\;
E_2 =   E_{2,3},\;
F_0 = z^{-1} E_{1,3},\;
F_1 =   E_{2,1},\;
F_2 =   E_{3,2},
\label{Cheva}
\end{equation}
$$
H_0 = c+ E_{3,3}-E_{1,1},\quad
H_1 = E_{1,1}-E_{2,2},\quad
H_2 = E_{2,2}-E_{3,3},
$$
where $E_{i,j}$ is the matrix unit
$E_{i,j}=(\delta_{ia}\delta_{jb})_{a,b=1}^3.$
The Cartan subalgebra of $\hat{\g}$ is defined as
$
\hat{\h}=\bigoplus_{i=0}^2\C H_i \oplus \C d.
$
We introduce the simple roots $\alpha_j$ and the fundamental weights
$\Lambda_j$
as the following linear functionals on the Cartan subalgebra $\hat{\h}$:
$$
\langle H_i, \alpha_j \rangle = a_{ij},\quad
\langle H_i, \Lambda_j \rangle = \delta_{ij}\quad (i=0,1,2),\quad
\langle d, \alpha_j \rangle   = \delta_{0j},\quad
\langle d, \Lambda_j \rangle   = 0
$$
for $j=0,1,2,$
where $(a_{ij})_{i=0}^3$ is the generalized Cartan matrix 
of type $A_2^{(1)}$ defined by
$$
\left[
  \begin{array}{@{\,}ccc@{\,}}
     2 & -1 & -1 \\
    -1 &  2 & -1 \\
    -1 & -1 &  2
  \end{array}
\right].
$$
We define a non degenerate symmetric bilinear 
form $(\,.\,|\,.\,)$ on $V={\hat{\h}}^*$ as follows:
$$
(\alpha_i|\alpha_j)=a_{ij},\quad
(\alpha_i|\Lambda_0)=\delta_{i0},\quad
(\Lambda_0|\Lambda_0)=0.
$$

We define simple reflections $s_i\,(i=0,1,2)$ by
$$
s_i(\lambda)=\lambda-\langle H_i, \lambda\rangle\,\alpha_i,
\quad \lambda\in V.
$$
They satisfy the fundamental relations
$$
s_i^2=1,\quad
s_is_{i+1}s_i=s_{i+1}s_is_{i+1}\quad(i=0,1,2),
$$
where the indices are understood as elements of $\Z/3\Z$.
Consider the group
\begin{equation}
  W = \langle s_0,s_1,s_2\rangle \subset {\mathrm{GL}}(V),
\label{Weylgroup}
\end{equation}
generated by the simple reflections. The group $W$ is
called the affine Weyl group of type $A_2^{(1)}.$

\section{Modified Yajima-Oikawa hierarchy}
\label{hier}
\setcounter{equation}{0}

In this section we introduce the modified Yajima-Oikawa 
hierarchy as generalized Drinfel'd-Sokolov reduction
associated to the loop algebra $
{L\g}=\sl_3\left(\C[z,z^{-1}]\right),
$ following \cite{gds1}.
Let us introduce the following derivation on $L{\g}$:
\beqn
  D = 4z\frac{\rd}{\rd z} - \diag(-1,0,1).
\label{def-d}
\eeqn
Set
$$
  L\g_j = \{A\in L\g\;|\;[D,A]=jA\}.
$$
Then we have a $\Z$-gradation $L\g=\oplus_{j}L\g_j$.
Note that
\begin{equation*}
 \deg(E_0)=-\deg(F_0)=2,\quad
 \deg(E_j)=-\deg(F_j)=1 \quad (j=1,2).
\end{equation*}
Consider the particular element
$$
\gamma= \left[
  \begin{array}{@{\,}ccc@{\,}}
    0 & 0 & 1\\
    0 & 0 & 0\\
    z & 0 & 0
  \end{array}
\right]
$$
and let $\s$ be the centralizer of $\gamma$ in $L\g$
$$
  \s = {\mathrm{Ker}} \left({\mathrm{ad}} \gamma\right)
     = \{ A \in L\g \; | \; [\gamma,A] = 0 \}.
$$
The subalgebra $\s$ is a maximal commutative subalgebra
in ${\g}$, which has the following basis:
$$
 \gamma_{4j+2} = z^{j} \gamma, \quad
 \gamma_{4j}   = z^j\diag(1,-2,1) \quad (j\in \Z).
$$
Then $\s$ is a graded subalgebra of $L\g$ with 
respect to the gradation. We have $\gamma_{2j} \in L\g_{2j}$.
The commutative subalgebra $\s$ is the image of a Heisenberg subalgebra in
$\hat{\g}$ 
associated with the conjugacy 
class $(2,1)$ (\cite{112}, see also \cite{tkl1} and \cite{Leid}).
We put $\b:=\oplus_{j \ge 0} L\g_j$. 

To introduce our hierarchy,
we begin with the differential operator
$$
  L:=\frac{\rd}{\rd x} - \gamma - Q,
$$
where $Q$ is an $x$-dependent element of $\b_{<2}.$
We set $\s^{\perp}:={\mathrm{Im}}\left({\mathrm{ad}\gamma}\right).$
It is clear that $\s^{\perp}=\oplus_{j}\s^{\perp}_{j},$ where
$\s^{\perp}_{j}:=\s^{\perp}\cap L\g_{j}.$
There is a unique formal series
$U=\sum_{j=1}^{\infty}U_{-j}\;(U_{-j}\in \s^{\perp}_{-j})$
such that the operator $L_0:=e^{{\mathrm{ad}}U}(L)$ has the form
$$
  L_0 = \frac{\rd}{\rd x} - \gamma - \sum_{j=0}^{\infty}h_{-2j},
 \quad  h_{-2j} \in \s_{-2j}.
$$
Moreover $U_{-j}$ and $h_{-2j}$ are polynomials in the 
components of $Q$ and their $x$ derivatives.
For any $j>0$ we set
$$
  B_{2j} = \left(e^{-{\mathrm{ad}}U}\gamma_{2j}\right)_{\ge 0}.
$$
The modified Yajima-Oikawa hierarchy is defined by the Lax equations
$$
  \frac{\rd L}{\rd t_{2j}} = [B_{2j},L] \quad (j=1,2,\ldots).
$$

We describe the above construction concretely. 
First we set
$$
 Q = \left[ \begin{array}{@{\,}ccc@{\,}}
               u_0 &  r  &  0  \\
                0  & u_1 &  q  \\
                0  &  0  & u_2
  \end{array} \right], 
\quad u_0 + u_1 + u_2 =0
$$
and solve for the first few terms of $U_j$ and $h_j$:
\begin{align*}
  U_{-1} =& -qE_{2,1} + rE_{3,2},
\\
  U_{-2} =& \frac{u_2-u_0}{4} (z^{-1}E_{1,3}-E_{3,1}),
\\
  U_{-3} =& \left[ \left(\frac{3u_0}{8} + \frac{3u_1}{2} - \frac{3u_2}{8} 
               - qr \right)r + r' \right] E_{1,2} \\
         & \quad + \left[ \left(\frac{7u_0}{8} - u_1 + \frac{u_2}{8} 
               + qr \right)q + q' \right] E_{2,3},
\\
  U_{-4} =& \left[ \frac{u_0'-u_2'}{8} + \frac{q'r+3qr'}{8}
            + \left(\frac{u_0}{16} - \frac{5u_1}{16} + \frac{u_2}{16}
               + \frac{5}{16}qr \right)qr \right] (E_{1,1}-E_{3,3}),
\end{align*}
\begin{align*}
  h_0 &=  \frac{qr - u_1}{2} \gamma_0,
\\
  h_{-2} &= \left[ \frac{u_0^2+u_2^2}{8} - \frac{u_0u_2}{4} 
            - \frac{q'r + 3qr'}{4} 
            - \left(\frac{u_0}{8} - \frac{5u_1}{8} + \frac{u_2}{8} 
                     + \frac{3}{8}qr \right) qr \right] \gamma_{-2}.
\end{align*}
Here $'$ means $\rd/\rd x$.
In fact, $h_0$ is a constant along 
all the flows and we can put $h_0 = 0$ (see \cite{gds1}).
So we fix
\beqn
 u_1 = qr
\label{u1qr}
\eeqn
from now on.
By using $U_j$'s and condition (\ref{u1qr}) we have
\beqn
B_2= \gamma_2 + \left[\begin{array}{@{\,}ccc@{\,}}
    u_0 &  r & 0 \\
     0  & u_1& q \\
     0  &  0 &u_2
  \end{array}
\right],
\label{B2}
\eeqn
\beqn
B_4 =\gamma_4
+
3\left[
  \begin{array}{@{\,}ccc@{\,}}
    -qr'+qru_2 & r'-ru_2       & 0        \\
    qz        & qr'-q'r+qru_1 & -q'-qu_0 \\
    -qrz      &        rz     & q'r+qru_0
  \end{array}
\right]
\label{B4}
\eeqn
The modified Yajima-Oikawa equation is obtained by 
the following zero-curvature condition:
\beqn
\frac{\rd B_2}{\rd t_4} = \frac{\rd B_4}{\rd t_2} - [B_2, B_4].
\label{zero-curv}
\eeqn
In fact this yields
the following system of differential equations:
\begin{align}
  q_t &+ 3\left(q''+q(-qr'+u_0'+qru_2+u_2^2)\right) = 0,  
\label{mYOq} \\
  r_t &- 3\left(r''-r(-q'r+u_2'-qru_0+u_0^2)\right) = 0,
\label{mYOr}
\end{align}%
\begin{equation}
  (u_0)_t = 3(-qr'+qru_2)',\quad
  (u_1)_t = 3(qr' - q'r + qru_1)', \quad
  (u_2)_t = 3(q'r+qru_0)'.
\label{mYOu}
\end{equation}
Here we identify $x$ and $t_2$, and put $t = t_4$.

\textbf{Remark}:
This system of equations is related to the Yajima-Oikawa 
equation \cite{yo1}:
\begin{align}
   &\Psi_t + 3\left(\Psi''+ u \Psi  \right) = 0,
\label{yopsi} \\
   &\Phi_t - 3\left(\Phi''+ u \Phi  \right) = 0,
\label{yophi} \\
   &u_t + 6(\Psi \Phi)' = 0.
\label{you}
\end{align}

The relation is established by the following map, which takes a
solution $q$, $r$, $u_j$ $(j=0,1,2)$ of (\ref{mYOq}) 
(\ref{mYOr}), (\ref{mYOu}) into a solution
$\Psi$, $\Phi$, $u$ of (\ref{yopsi}), (\ref{yophi}), (\ref{you})
and is an analog of the Miura map in
the case of KdV and mKdV equations:
$$
 \Psi = -q'-qu_0, \quad
 \Phi = r'-ru_2,  \quad
  -u  = u_0^2 + u_2^2 + u_0u_2 + u_0' + qr'.
$$

\section{Similarity reduction}
\label{Similarity}
\setcounter{equation}{0}

In this section
we consider a self-similarity condition
on the solutions of 
the modified Yajima-Oikawa equation
(\ref{mYOq}), (\ref{mYOr}), (\ref{mYOu}).
These are the main object of this paper.
A solution $q(x,t)$, $r(x,t)$, $u_j(x,t)$ $(j=0,1,2)$ is said to 
be self-similar if
\beqn
q(\lambda^2 x,\lambda^4 t)=\lambda^{-1}q(x,t),\quad
r(\lambda^2 x,\lambda^4 t)=\lambda^{-1}r(x,t),\quad
u_j(\lambda^2 x,\lambda^4 t)=\lambda^{-2}u_j(x,t).\label{similar}
\eeqn
Here we count a degree of variables by
$\deg x = \deg t_2 = -2$, $\deg t = \deg t_4 = -4$.
Note that such functions are uniquely 
determined by its values at fixed $t$, 
say at $t=1/4$.
Differentiating (\ref{similar}) with respect to $\lambda$ at $\lambda=1$,
we obtain the Euler equations
$$
2x\frac{\rd q}{\rd x}+4t\frac{\rd q}{\rd t}=-q,\quad
2x\frac{\rd r}{\rd x}+4t\frac{\rd r}{\rd t}=-r,\quad
2x\frac{\rd u_j}{\rd x}+4t\frac{\rd u_j}{\rd t}=-2u_j.
$$
At $t=1/4$ these identities become
$$
\frac{\rd q}{\rd t}=-2\frac{\rd (xq)}{\rd x}+q,\quad
\frac{\rd r}{\rd t}=-2\frac{\rd (xr)}{\rd x}+r,\quad
\frac{\rd u_j}{\rd t}=-2\frac{\rd (xu_j)}{\rd x}.
$$
This can be written in the matrix form
$$
 \frac{\rd B_2}{\rd t} = -2 \frac{\rd(xB_2)}{\rd x}+[D,B_2],
$$
where $D$ is the derivation defined in (\ref{def-d}).
Substituting this last identity
into the zero-curvature equation (\ref{zero-curv}),
we obtain
\beqn
\frac{\rd M}{\rd x}=
\left[
  4z\frac{\rd}{\rd z}-M,B_2
\right],
\eeqn
where we set
\begin{equation}
  M =\left[\begin{array}{@{\,}ccc@{\,}}
              \ep_1 &  f_1   & g      \\
                0   & \ep_2  & f_2    \\
                0   &   0    & \ep_3
  \end{array}\right] + z\left[\begin{array}{ccc}
                 1  &  0  &  0   \\
                3q  & -2  &  0   \\
                f_0 & 3r  & 1
  \end{array}\right]
 := \diag (-1,0,1) + 2x B_2 + B_4.
\label{M} 
\end{equation}
The correspondence of variables is given as follows:
v\begin{align}
\ep_1 &= -1+2xu_0-3q(r'-ru_2),   \label{ep1}\\
\ep_2  &= 2xu_1+3(qr'-q'r+qru_1),  \label{ep2} \\
\ep_3 &= 1+2xu_2+3r(q'+qu_0)      \label{ep3}
\end{align}
and $g=2x$, 
\beqn
f_0=2x-3qr,\quad
f_1=2xr+3(r'-ru_2),\quad
f_2=2xq-3(q'+qu_0).
\eeqn
Here we regard the 
variables $q = q(x,1/4)$, $r = r(x,1/4)$, $u_j = u_j(x,1/4)$ 
$(j=0,1,2)$ are functions only in $x$.
Note that the definition of $M$ has a freedom
of adding a constant diagonal matrix and here 
we normalize
\begin{equation}
 \ep_1 + \ep_2 + \ep_3 = 0.
\label{epsilon}
\end{equation}

\section{Lax pair formalism}
\label{Laxpair}
\setcounter{equation}{0}
Consider the following system of linear differential equations
for the column vector $\psv={}^t(\psi_1,\psi_2,\psi_3)$ of three unknown
functions $\psi_i=\psi_i(z,x)\;(i=1,2,3)$ :
\beqn
 4z\frac{\rd}{\rd z}\psv = M\psv,\quad
 \frac{\rd}{\rd x}\psv=B\psv.
\label{eq:wave}
\eeqn
We assume that the matrix $M$ is (\ref{M}) 
and $B = B_2$ (\ref{B2})
where the variables $\ep_j,f_j,u_j,q,r$ and $g$ are
functions in $x$.
Then the compatibility condition of system (\ref{eq:wave})
\beqn
\left[4z\frac{\rd}{\rd z}-M,\frac{\rd}{\rd x}-B\right]=0
\label{lax}
\eeqn
is 
equivalent to the relations
\beqn
\begin{array}{ll}
  \ep_1' = \ep_2' = \ep_3' = 0,   & g=f_0+3qr,       \\
    f_0' = f_0(u_2-u_0)-(\ep_3-\ep_1-4), & g'
    =g(u_0-u_2)-qf_1+rf_2-\ep_1+\ep_3,        \\
      f_1'=f_1(u_0-u_1)-r(\ep_1-\ep_2),    & 3q'=3q(u_1-u_0)+qf_0-f_2, 
      \\
      f_2'=f_2(u_1-u_2)-q(\ep_2-\ep_3),    &
      3r'=3r(u_2-u_1)-rf_0+f_1. 
\label{compati}
\end{array}
\eeqn
If we forget the 
relation \eqref{M} of $M$ and $B_1$, $B_2$ and 
start from the Lax equation \eqref{compati},
we can recover some of the relations of variables.
For instance, differentiating 
both-hand side of $g = f_0 + 3qr$ and 
eliminate the variables except $g'$ by means of (\ref{compati}),
we get $g'=2$ and therefore assume 
$$ 
  g = 2x.
$$

In what follows we shall impose the following constraint 
on the variables:
\beqn
  u_0 + u_1 + u_2 = 0,  \quad  u_1=qr.
\label{constr}
\eeqn
The joint system (\ref{lax}) and (\ref{constr}) is the main object that we 
investigate in this paper.
Using system (\ref{compati}) together with the constraint,
we can derive the following equation:
\beqn
    2gu_0=qf_1-rf_2-gqr-\ep_3+\ep_1+2.
\label{elim}
\eeqn
After the elimination of the variables $f_0,u_0,u_1,u_2$ by
(\ref{compati}), (\ref{constr}) and (\ref{elim}), we
obtain a system of ODE for the unknown functions $f_1,f_2,q,r$
with the parameters $\ep_1,\ep_2,\ep_3.$
We can obtain the set of explicit formulae of $f_1',f_2',q',r'$ in
terms of $f_1,f_2,q,r$ and $g$, and the results are
\begin{align}
    f_1' &= \frac{f_1}{2g}(f_1q - f_2r) - \frac{3}{2}f_1qr 
            + (\ep_1 - \ep_3)\frac{f_1}{2g} - (\ep_1-\ep_2)r 
            + \frac{f_1}{g},  
\label{f1}  \\
    f_2' &= \frac{f_2}{2g}(f_1q - f_2r) + \frac{3}{2}f_2qr 
            + (\ep_1 - \ep_3)\frac{f_2}{2g} - (\ep_2 - \ep_3)q 
            + \frac{f_2}{g}, 
\label{f2}\\
    q'  &=  - \frac{q}{2g}(f_1q - f_2r) + \frac{q^2r}{2} 
            - (\ep_1 - \ep_3)\frac{q}{2g} + \frac{gq - f_2}{3}
            - \frac{q}{g}, 
\label{q}\\
    r'  &=  - \frac{r}{2g}(f_1q - f_2r) - \frac{qr^2}{2} 
            - (\ep_1 - \ep_2)\frac{r}{2g} - \frac{gr - f_1}{3}
            - \frac{r}{g}. 
\label{r}
\end{align}
In Sect.\ref{Hamil} we present the system of ODE in 
the Hamiltonian form.

{\bf{Remark.}}
Using (\ref{elim}) and (\ref{compati}),
we can also derive the following differential equation:
\beqn
  gu_0'=(\ep_2-\ep_3)qr+\frac{f_2}{3}(rf_0-f_1)-2u_0.
\eeqn

\section{B\"acklund transformations}
\label{Backlund}
\setcounter{equation}{0}

Let us pass to the investigation of a group of B\"acklund
transformations.
For this purpose, it is convenient to introduce the following set of
parameters:
\begin{equation}
  \alpha_0=\ep_3-\ep_1-4,  \quad
  \alpha_1=\ep_1-\ep_2,    \quad
  \alpha_2=\ep_2-\ep_3.
\label{parameters}
\end{equation}
They are identified with the simple roots
of the affine root system of type $A_2^{(1)}$. 

We define the B\"acklund transformations for the system
by considering the gauge transformations of the linear 
system (\ref{eq:wave})
\beqn
    s_i\psv=G_i\psv\quad (i=0,1,2).
\label{spsv}
\eeqn
The matrices $G_i$ are given as follows:
\beqn
    G_i=1+\frac{\alpha_i}{f_i}F_i\quad(i=0,1,2),
\label{Gi}
\eeqn
where $F_0,F_1,F_2$ are Chevalley generators \eqref{Cheva} of 
the loop algebra $\sl_3(\C[z,z^{-1}])$.
The compatibility condition of (\ref{eq:wave}) and (\ref{spsv}) is
\beqn
  s_i(M) = G_iMG_i^{-1} + 4z\frac{\rd G_i}{\rd z}G_i^{-1},
\quad
  s_i(B) = G_iBG_i^{-1} + \frac{\rd G_i}{\rd x}G_i^{-1}.
\label{compati_G}
\eeqn
On the components of the matrices $M,B$, the actions of $s_i(i=0,1,2)$
are given explicitly as in the following tables:

\vspace{10pt}

\begin{center}
\begin{tabular}{c|ccc|ccc}
      & $f_0$ & $f_1$ & $f_2$  & $g$ & $q$ & $r$ \\ \hline
$s_0$ & $f_0$ & $f_1+3r\frac{\al_0}{f_0}$ & $f_2-3q\frac{\al_0}{f_0}$  
      & $g$ & $q$ & $r$  \\
$s_1$ & $f_0-3r\frac{\al_1}{f_1}$ & $f_1$ & $f_2+g\frac{\al_1}{f_1}$ &
         $g$ & $q+ \frac{\al_1}{f_1}$ & $r$  \\
$s_2$ & $f_0+3q\frac{\al_2}{f_2}$ & $f_1-g\frac{\al_2}{f_2}$ & $f_2$ &
         $g$ & $q$ & $r-\frac{\al_2}{f_2}$   \\
\end{tabular}
\end{center}

\vspace{10pt}

\begin{center}
\begin{tabular}{c|ccc|ccc}
      & $\al_0$ & $\al_1$ & $\al_2$ & $u_0$ & $u_1$ & $u_2$ \\ \hline
$s_0$ & $-\al_0$ & $\al_1+\al_0$ & $\al_2 + \al_0$  
      & $u_0+\frac{\alpha_0}{f_0}$ & $u_1$ & $u_2-\frac{\alpha_0}{f_0}$\\
$s_1$ & $\al_0 + \al_1$ & $-\al_1$ & $\al_2 + \al_1$ 
      & $u_0-r\frac{\alpha_1}{f_1}$ & $u_1+r\frac{\alpha_1}{f_1}$ & $u_2$\\
$s_2$ & $\al_0 + \al_2$ & $\al_1 + \al_2$ & $-\al_2$ 
      & $u_0$ & $u_1-q\frac{\alpha_2}{f_2}$ & $u_2+q\frac{\alpha_2}{f_2}$\\
\end{tabular}
\end{center}

\vspace{10pt}

The automorphisms $s_i(i=0,1,2)$ generate a group of B\"acklund
transformations for our differential system.
To state this fact clearly,
it is convenient to introduce the field
\beqn
  K = \C(\alpha_0,\alpha_1,\alpha_2,f_0,f_1,f_2,g,q,r,u_0,u_1,u_2),
\label{fieldK}
\eeqn
where the generators satisfy the following algebraic relations:
$$
  \alpha_0+\alpha_1+\alpha_2=-4,\quad
  f_0 = g - 3qr,\quad
  u_0 + u_1 + u_2=0,\quad
  u_1=qr,
$$
$$
    2gu_0=qf_1-rf_2-gqr-\ep_3+\ep_1+2.
$$
We have the automorphisms $s_i\;(i=0,1,2)$ of the
field $K$ defined by the above table.
Note that the field $K$ is thought to be a
differential field with the derivation $':K \to K$ defined 
by (\ref{compati}).

\begin{thm}\label{thm:Weyl}
   The automorphism $s_0,s_1,s_2$ of $K$ define a representation of 
   the affine Weyl group $W$ $\eqref{Weylgroup}$ on the field $K$
    such that the action of the each element $w\in W$ commutes with
   the derivation of the differential field $K$.
\end{thm}

Theorem \ref{thm:Weyl} is proved by straightforward computations.
Note that the independent variable $x=g/2$ is fixed under 
the action of $W.$

\section{Hamiltonian structure}
\label{Hamil}
\setcounter{equation}{0}

We shall equip $K$ (\ref{fieldK}) with the Poisson algebra
structure $\{ \;,\; \}: K \times K \to K$ defined as follows:
$$
\begin{array}{c|cccc}
   \{\ ,\ \} & f_1 & f_2 & q & r  \\  \hline
      f_1    &  0  &  g  & 1 & 0  \\
      f_2    & -g  &  0  & 0 & -1 \\
       q     & -1  &  0  & 0 & 0  \\
       r     &  0  &  1  & 0 & 0  
\end{array}
$$
That is, $\{ f_1, f_2 \} = g$ and so on.
Note that the Poisson structure comes from the Lie algebra structure
of $\hat{\g}$ (see \cite{Noumi} for an exposition).
We can describe the action
of $s_i$ $(i=0,1,2)$ on the 
generators $f=f_j,u_j,q,r,g$ $(j=0,1,2)$ of $K$
by
$$
 s_i(f) = f + \frac{\alpha_i}{f_i} \{f_i, f \}.
$$

We introduce the function $h$ by
\begin{align*}
  h :=&  \frac{1}{2}(f_1q^2r + f_2qr^2)
       - \frac{1}{4g}(f_1^2q^2 + f_2^2r^2 + q^2r^2g^2)  
       + \left( \frac{qr}{2g} - \frac{1}{3} \right) f_1f_2  \\
     & + \left( \frac{g}{3} - \frac{\al_1+\al_2}{2g} \right) f_1q 
       + \left( \frac{g}{3} + \frac{\al_1+\al_2}{2g} \right) f_2r 
       - \left( \frac{g}{3} - \frac{\al_1-\al_2}{2g} \right) qrg.
\end{align*}
Then the differential system (\ref{f1})--(\ref{r}) can be expressed
\begin{align}
  f_1' &= \{h, f_1\} + \frac{f_1}{g}, \quad
   q'   = \{h, q \}  - \frac{q}{g},    
\label{poisson} \\
  f_2' &= \{h, f_2\} + \frac{f_2}{g}, \quad
   r'   = \{h, r \}  - \frac{r}{g}.
\nonumber
\end{align}
Let us introduce the variables
$$
  p_1 = f_1,            \quad
  q_1 = q,              \quad
  p_2 = \frac{f_2}{g}-q,\quad
  q_2 = -gr.
$$
It is easy to show that
$$
     \{p_i,q_j\}=\delta_{ij},\quad
     \{p_i,p_j\}=\{q_i,q_j\}=0 \quad(i,j=1,2).
$$

\begin{thm}\label{thm:Hamil}
       Let $H$ be the function defined as
\begin{align*}
  xH = & - \frac{1}{4}p_1p_2q_1q_2
         - \frac{1}{8}\left( p_1^2q_1^2+p_2^2q_2^2 \right)
         - \frac{1}{2}p_1q_1^2q_2  \\
       & - \frac{1}{4}\left( \al_1 + \al_2 + 2 \right)p_1q_1
         - \frac{1}{4}\left( \al_1 + \al_2 - 2 \right)p_2q_2
         - \frac{\al_1}{2} q_1q_2
         - \frac{2x^2}{3} (q_2 + p_1)p_2
\end{align*}
Then the system of ODEs $(\ref{f1})$, $(\ref{f2})$,
$(\ref{q})$, $(\ref{r})$ is equivalent to the
Hamiltonian system
\beqn
       \frac{d q_1}{dx}=\frac{\rd H}{\rd p_1},\quad
       \frac{d q_2}{dx}=\frac{\rd H}{\rd p_2},\quad
       \frac{d p_1}{dx}=-\frac{\rd H}{\rd q_1},\quad
       \frac{d p_2}{dx}=-\frac{\rd H}{\rd q_2}.
\label{poisson2}
\eeqn
\end{thm}

\begin{proof}
We define
$$
  H = h - \frac{f_1q + f_2r}{g} + qr
$$
and rewrite this in the coordinate $p_j$, $q_j$ $(j=1,2)$.
Then the equations (\ref{poisson}) can be 
expressed as (\ref{poisson2}).
\qed 
\end{proof}

The behavior of the Hamiltonian under the B\"acklund transformations
is given by the simple formulae
$$
   s_0(\tilde{H}) = \tilde{H} + 6qr\frac{\alpha_0}{f_0},\quad
   s_j(\tilde{H}) = \tilde{H}  \;(j=1,2),
$$
where we set $\tilde{H} = xH + a$ with the correction term
$$
   a = \frac{1}{24}(\alpha_1 - \alpha_2)(\alpha_1 - \alpha_2 - 4).
$$

\section{Reduction to the fifth Painlev\'e equation}
\label{Painleve}
\setcounter{equation}{0}

In this section, we show the system (\ref{compati}) is
equivalent to a two-parameter family of the fifth Painlev\'e equation.
By linear change of the independent variable, we ensure the
normalization
\begin{equation}
  f_0 + \frac{f_1}{r} + \frac{f_2}{q} 
      + 3\left( \frac{q'}{q} - \frac{r'}{r} \right) = 3g = 6x
\label{6x}
\end{equation}
holds.
After the elimination of $u_0$ and $u_2$, we have
\begin{align}
     f_0'  &= -\frac{f_0}{3} \left( \frac{f_1}{r} - \frac{f_2}{q} \right)
                 + \frac{f_0u_1'}{u_1} - \alpha_0
\label{f0} 
\\
 \left(\frac{f_1}{r} \right)' 
    &= -\frac{f_1}{3r} 
       \left(\frac{f_2}{q} - f_0 + \frac{3u_1'}{u_1} \right)
      - \alpha_1, 
\label{f1r} \\
 \left( \frac{f_2}{q} \right)'
    &= -\frac{f_2}{3q} 
       \left(f_0 - \frac{f_1}{r} + \frac{3u_1'}{u_1}\right)
      - \alpha_2.
\label{f2q}
\end{align}
Here we introduce a new variable
$$
  y := - \frac{f_0}{3u_1}.
$$
Notice the relations
\beqn
 y - 1 = -\frac{2x}{3u_1}, \quad
 \frac{y'}{y-1} = \frac{1}{x} - \frac{u_1'}{u_1}
\label{utoy}
\eeqn
holds by $f_0 = g - 3qr = 2x - 3u_1$.
Then we rewrite (\ref{f0}) as
\begin{equation}
   y'  = -\frac{y}{3} \left( \frac{f_1}{r} - \frac{f_2}{q} \right)
           + \frac{\alpha_0}{3u_1}, 
\label{y1}
\end{equation}
After differentiating (\ref{y1}), elimination of the 
variables $f_1$, $f_2$, $q$, $r$, $u_1$ by
(\ref{6x}), \eqref{f1r}, (\ref{f2q}), (\ref{utoy}), (\ref{y1}) 
and the definition of the constant $\ep_2$ (\ref{ep2})
leads to the following equation of $y$:
\begin{align}
   y'' =& \left(\frac{1}{2y} 
          + \frac{1}{y-1} \right) (y')^2
          - \frac{y'}{x}
          + \frac{(y-1)^2}{8x^2} \left(\ep_2^2y
          - \frac{\al_0^2}{y} \right) \nonumber \\
        & \quad - \frac{2x^2y}{9} - \frac{4x^2y}{9(y-1)}  
          - \frac{(\al_2-\al_1)y}{3} 
          + \frac{\ep_2y}{3}.
\label{y''3}
\end{align}
We put $\xi = x^2$, then the equation (\ref{y''3}) can
be brought into the fifth Painlev\'e equation
\begin{align*}
   y_{\xi\xi} =& \left(\frac{1}{2y} 
          + \frac{1}{y-1} \right) (y_{\xi})^2
          - \frac{1}{\xi}y_{\xi}
          + \frac{(y-1)^2}{\xi^2} \left(Ay + \frac{B}{y} \right) 
          + \frac{C}{\xi}y
          - \frac{y(y+1)}{18(y-1)},
\end{align*}
where
$$
 A = \frac{\ep_2^2}{32}, \quad
 B = - \frac{\al_0^2}{32}, \quad
 C = -\frac{\ep_2}{6}.
$$
Note that $\ep_2 = (\alpha_2-\alpha_1)/3$ holds 
by \eqref{epsilon} and \eqref{parameters}.

\section{$\tau$-functions}
\label{taufunctions}
\setcounter{equation}{0}

We introduce the $\tau$-functions $\tau_0$,$\tau_1$,$\tau_2$,
$\sigma_1$ and $\sigma_2$ to
be the dependent variables satisfying the following equations:
\beqn
 \frac{f_1}{r} = 2x +3\left(
       \frac{\sigma_2'}{\sigma_2}
       -\frac{\tau_0'}{\tau_0} \right),\quad
 \frac{f_2}{q} = 2x -3\left(
        \frac{\sigma_1'}{\sigma_1}
       -\frac{\tau_0'}{\tau_0} \right),\quad
 q = -\frac{\sigma_1}{\tau_1}, \quad
 r = \frac{\sigma_2}{\tau_2}.
\label{defoftau}
\eeqn
To fix the freedom of overall multiplication by a function in 
the defining equation (\ref{defoftau}) for
$\tau_0$, $\tau_1$, $\tau_2$, $\sigma_1$ and $\sigma_2$, we 
impose the equation
\begin{align}
    \left(\log\tau_0^2\tau_1^2\tau_2^2\sigma_1\sigma_2\right)''
 &+ u_0^2 + u_2^2 
       + \left( u_0 - \frac{f_1}{3r} + \frac{2x}{3} \right)^2
       + \left( u_2 - \frac{f_2}{3q} + \frac{2x}{3} \right)^2  
\nonumber \\
 &- \frac{2x}{9}\left(4x - \frac{f_1}{r} - \frac{f_2}{q} \right)
 -\frac{\alpha_1-\alpha_2}{9}
    =0.
\label{NorTau}
\end{align}

The differential equations for the variables $q$ and $r$ in 
the system (\ref{compati}) lead to
\beqn
   u_0 = \frac{\tau_1'}{\tau_1}-\frac{\tau_0'}{\tau_0},
\quad
   u_2 = \frac{\tau_0'}{\tau_0}-\frac{\tau_2'}{\tau_2}
\label{u2}
\eeqn
respectively. Here we have used the relations 
$$
 u_1 = qr =  -\frac{\sigma_1\sigma_2}{\tau_1\tau_2}, \quad
 f_0 = 2x - 3qr = 2x + 3\frac{\sigma_1\sigma_2}{\tau_1\tau_2}.
$$
If the equations (\ref{u2}) are satisfied, we have
\beqn
  u_1 = \frac{\tau_2'}{\tau_2}-\frac{\tau_1'}{\tau_1},
\label{u1}
\eeqn
by $u_0+u_1+u_2=0$ and therefore have the following formula of 
the variable $f_0$ in terms of the $\tau$-functions:
\beqn
  f_0 = 2x +3\left(
       \frac{\tau_1'}{\tau_1}
       -\frac{\tau_2'}{\tau_2}
     \right).
\eeqn
Let $D_x$ and $D_x^2$ be Hirota's bilinear operators:
$$
  D_x F\cdot G := F'G - FG', \quad
  D_x^2 F\cdot G := F''G - 2F'G' + FG''.
$$
In this notation, the relation $u_1=qr$, for
example, can be written in
\beqn
  D_x\tau_1\cdot\tau_2 = \sigma_1\sigma_2.
\label{D1}
\eeqn

We introduce a system of bilinear equations that
leads to our differential system (\ref{compati}).
\begin{thm}\label{thm:bilinear}
  Let $\tau_0,\tau_1,\tau_2,\sigma_1,\sigma_2$ be a set of functions 
that satisfies the following system of
Hirota bilinear equations$:$
\begin{align}
  &\left(3D_x^2 - 2xD_x + \frac{1}{6}(\alpha_0-4\alpha_1-2)\right)
         \tau_0\cdot\tau_1 = 0,     \label{2ndbin1}\\
  &\left(3D_x^2 - 2xD_x -\frac{1}{6}(\alpha_0-4\alpha_2-2)\right)
         \tau_2\cdot\tau_0 = 0,     \label{2ndbin2}\\
  &\left(3D_x^2 - 2xD_x +\frac{1}{6}(\alpha_1-\alpha_2+6)\right)
         \tau_1\cdot\sigma_2 = 0,   \label{2ndbin3}\\
  &\left(3D_x^2-2xD_x + \frac{1}{6}(\alpha_1-\alpha_2-6)\right)
         \sigma_1\cdot\tau_2  = 0,    \label{2ndbin4}
\end{align}
together with $(\ref{D1})$. If we define the 
functions $f_0$, $f_1$, $f_2$, $q$, $r$, $u_0$, $u_1$ and $u_2$ 
by the formulae $(\ref{defoftau})$, $(\ref{u2})$, $(\ref{u1})$ then 
this set of functions satisfies our ODE system $(\ref{compati})$ together 
with algebraic equations $(\ref{constr})$.
\end{thm}

\begin{proof} 
We can verify that the differential equations 
for $q$ and $r$ are satisfied if we assume the 
existence of the $\tau$-functions such that
equations (\ref{defoftau}), (\ref{u2}) holds.
The differential equations for $f_0$ is written as
\beqn
  3(g_1''-g_2'') + 2 = (3(g_1'-g_2')+2x)(2g_0'-g_1'-g_2') - \alpha_0,
\label{tau_f0}
\eeqn
where $g_j=\log \tau_j$, $(j=0,1,2)$. This equation is obtained 
if we subtract (\ref{2ndbin1}) from (\ref{2ndbin2}).
The differential equations for $f_1$ and $f_2$ can be rewritten as
\beqn
       \left(\frac{f_1}{r}\right)'
=\frac{f_1}{r}\left(u_0-u_1-\frac{r'}{r}\right)-\alpha_1\quad
       \left(\frac{f_2}{q}\right)'
=\frac{f_2}{q}\left(u_1-u_2-\frac{q'}{q}\right)-\alpha_2,
\eeqn
respectively.
In terms of the $\tau$-functions, these equations read
\begin{align}
  3(h_2''-g_0'') + 2 &= 
       (3(h_2'-g_0')+2x)(2g_1'-g_0'-h_2') - \alpha_1,
\label{f1_tau}\\
  3(g_0''-h_1'') + 2 &= 
       (3(g_0'-h_1')+2x)(2g_2'-g_0'-g_1') - \alpha_2,
\label{f2_tau}
\end{align}
where $h_1=\log \sigma_1, h_2=\log \sigma_2.$
In fact, from (\ref{2ndbin1}) and (\ref{2ndbin3})
we can eliminate $g_1''$ to obtain (\ref{f1_tau}). 
In the similar way from (\ref{2ndbin2}) and (\ref{2ndbin4}), 
we can eliminate $g_2''$ to obtain (\ref{f2_tau}).
\qed \end{proof}

We remark that the normalization of $\tau$-functions (\ref{NorTau})
is obtained by taking the sum of four equations in this theorem.

\section{Jacobi-Trudi type formula}
\label{Jacobi}
\setcounter{equation}{0}

In this section we lift the action of $W$ to 
the $\tau$-functions.
Consider the field 
extension $\widetilde{K}=K(\tau_0,\tau_1,\tau_2,\sigma_1,\sigma_2)$.
Then we can prove the next Theorem 
by a direct computation. 
\begin{thm}\label{thm:tauaut}
We extend each automorphism $s_i$ of $K$ to an automorphism 
of the field $\widetilde{K}=K(\tau_0,\tau_1,\tau_2,\sigma_1,\sigma_2)$
by the formulae
$s_i(\tau_j)=\tau_j\;(i\ne j),\;
 s_i(\sigma_k)=\sigma_k\;(i\ne k)$ and 
\begin{align}
s_0(\tau_0) &= f_0\frac{\tau_{2}\tau_{1}}{\tau_0},\quad
s_1(\tau_1) = f_1\frac{\tau_{0}\tau_{2}}{\tau_1}, \quad
s_1(\sigma_1)=-(f_1 q+\alpha_1)\frac{\tau_0\tau_2}{\tau_1},
\label{s1tau} \\
s_2(\tau_2) &= f_2\frac{\tau_{1}\tau_{0}}{\tau_2}, \quad
s_2(\sigma_2)=(f_2 r-\alpha_2)\frac{\tau_1\tau_0}{\tau_2}.
\label{s2tau}
\end{align}
Then these automorphisms define a representation 
of $W$ on $\widetilde{K}$.
\end{thm}

Following \cite{affine}, we will 
describe the Weyl group orbit of the $\tau$-functions
(see also \cite{Noumi}). 
For any $w\in W$ and $k=0,1,2$, there exists a rational
function $\phi_w^{(k)}\in K$ such that
\beqn
w(\tau_k)
=\phi_w^{(k)}\prod_{i=0,1,2}
\tau_i^{\left(\alpha_i|w(\Lambda_k)\right)}.\label{wtau}
\eeqn
We shall give an expression of $\phi_w^{(k)}$ in terms
of the Jacobi-Trudi type determinant.

A subset $M$ of $\Z$ is called a Maya diagram
if $M\cap \Z_{\ge 0}$ and $M^c\cap \Z_{<0}$ are
finite sets.
We define an integer
$$
c(M):=\sharp\left(M\cap \Z_{\ge 0}\right)
-\sharp\left(M^c\cap \Z_{<0}\right)
$$
called the charge of $M.$
If $c(M)=r$, we can express $M$ as $\{i_k|k<r\}$ by using
an strictly increasing sequence $i_k\,(k<r)$ such
that $i_k=k$ for $k\ll r.$
Then we associate
a partition $\lambda=(\lambda_1,\lambda_2,\ldots)$
given by
$$
 \lambda_j=i_{r-j+1}-(r-j+1), \quad (j=1,2,\ldots).
$$
The Weyl group $W=\langle s_0,s_1,s_2\rangle$ can be
realized as a subgroup of
the group of bijections $w:\Z\rightarrow\Z$ by setting
$$
  s_k=\prod_{j\in \Z}\sigma_{3j+k-1}\quad (k=0,1,2),
$$
where $\sigma_i\;(i\in\Z)$ is
the adjacent transposition $(i,i+1).$
For a Maya diagram $M$ and $w\in W$, we see that $w(M)\subset \Z$ is 
also a Maya diagram of the same charge.

For any $w\in W$ and $k=0,1,2,$
let $\lambda=(\lambda_1,\ldots,\lambda_r)$ be
the partition corresponding to the Maya diagram $M=w(\Z_{<k}).$
We set
$$
  N_{\lambda}^{(k)}=
      \prod_{\substack{i<j \\ i\in M^c, j\in M}}(\ep_i-\ep_j),
$$
where we impose the relation $\ep_i-\ep_{i+3}=-4\;(i\in \Z),$
so we have $N_{\lambda}^{(k)}\in \C[\alpha_0,\alpha_1,\alpha_2].$
We can apply the following formula
due to Y.Yamada \cite{yamada}:
\beqn
   \phi_w(\Lambda_k) = N_\lambda^{(k)} 
       \det\left(g_{\lambda_j-j+i}^{(k-i+1)}\right)_{1\le i,j\le r}.
\label{jacobi-trudi}
\eeqn
Here $g_p^{(k)}$ $(k \in \Z/3\Z, p \in \Z_{>0})$ are the 
determinant of $p \times p$ matrix described as follows.
First we define $g_p^{(0)}$ by
$$
 g_p^{(0)} :=  \frac{1}{N_p^{(0)}}
    \left|\begin{array}{@{\,}cccccc@{\,}}
         f_{00}  & f_{01} & f_{02} &         &        &   0    \\
       \beta_1 & f_{11}    & f_{12} &  f_{13} &        &        \\
               & \ddots & \ddots & \ddots  & \ddots &        \\
               &        & \ddots & \ddots  & \ddots & f_{p-3,p-1}\\
               &        &        &\beta_{p-2}& f_{p-2,p-2}& f_{p-2,p-1}\\
          0    &        &        &          &\beta_{p-1} & f_{p-1,p-1} 
            \end{array} \right|,
$$
where the components are
\begin{align*}
 f_{i,i}   &= f_i \quad (f_{i+3} = f_i),  \\
 f_{i,i+1} &= g \, (i \equiv 1), \quad 
             3q \, (i \equiv 2), \quad
             3r \, (i \equiv 0), \\
 f_{i,i+2} &= 1 \, (i \equiv 1, 0), \quad 
             -2 \, (i \equiv 2),
\end{align*}
and
$
  \beta_j = \sum_{i=j}^{p-1}\al_i = \ep_j-\ep_p.
$
Then we put $g_p^{(1)} = \pi(g_p^{(0)})$ and 
$g_p^{(2)} = \pi^2(g_p^{(0)})$ by the automorphism $\pi$:
$$ 
  \pi(f_{ij}) = f_{i+1,j+1}, \quad \pi(\ep_j) = \ep_{j+1}.
$$

The formula (\ref{jacobi-trudi}) is valid since the 
action of $W=\langle s_0,s_1,s_2\rangle$ in our
setting is reduced from the action of $A_\infty$ (cf. \cite{Noumi}):
$$
 s_i(\al_i) = - \al_i,                       \quad
 s_i(\al_{i \pm 1}) = \al_{i \pm 1} + \al_i, \quad
 s_i(\al_j) = \al_j \; (j \ne i, i \pm 1),
$$
where $\al_j := \ep_j - \ep_{j+1}$ $(j \in \Z)$ and
$$
 s_k(f_{i,j}) = f_{i,j} 
                 + (\delta_{k+1,i}f_{k,j} - \delta_{j,k}f_{i,k+1})
                  \frac{\al_k}{f_k}.
$$

\section{Differential field of $\tau$-functions}
\label{WeylTau}
\setcounter{equation}{0}

In this section we give supplementary discussions
on the affine Weyl group action.
In particular,
we consider a differential field of $\tau$-functions
that naturally contains the fields $K$ and $\widetilde{K}$.
The field $\widehat{F}$ we consider can be presented as
\beqn
\C(\alpha_0, \alpha_1, \alpha_2,x; \tau_0,\tau_1,\tau_2,
\sigma_1,\sigma_2,\tau_0',\tau_1',\tau_2',\sigma_1',\sigma_2')
\label{Fhat}
\eeqn
with some relations discussed below. Then
the set of bilinear equations in 
Theorem \ref{thm:bilinear} makes $\widehat{F}$
into the differential field.
To show some basic facts on $\widehat{F}$, 
we introduce some intermediate fields. 

Let $F$ denote the extended field 
of $\C(\alpha_0, \alpha_1, \alpha_2,x)$ obtained by 
adjoining the variables $g_0'$, $g_1'$, $g_2'$,
$h_1'$, $h_2'$ with the following relations:
\begin{align}
 3(g_0'-2h_2'+h_1')(g_1'-g_2')
&+ 2x(g_0'-2g_1'+g_2') + \alpha_1 + 1 = 0,
\label{rel1}\\
 3(h_2'-2h_1'+g_0')(g_1'-g_2')
&+ 2x(g_1'-2g_2'+g_0') + \alpha_2 + 1 = 0
\label{rel2}.
\end{align}
As in the proof of Theorem \ref{thm:bilinear}, we will identify $g_j'$ with
$(\log \tau_j)'$ and 
$h_1',h_2'$ with $(\log \sigma_1)',$ $(\log \sigma_2)'$
respectively. Note that the relations \eqref{rel1}, \eqref{rel2}
correspond to (\ref{ep1}), (\ref{ep2}), (\ref{ep3}).
It is easy to see $F=\C(\alpha_0, \alpha_1, \alpha_2,x)(g_0',g_1',g_2')$, 
and $g_0'$, $g_1'$, $g_2'$ are algebraically independent 
over $\C(\alpha_0, \alpha_1, \alpha_2,x)$.
So if we fix $g_j''\in F\,(j=0,1,2)$ in an arbitrary way,
then we have a derivation on $F.$
Now we want to introduce a derivation on $F$ in such a way that is 
consistent with the bilinear equations.
Actually we can prove the following 
lemma by lengthy but straightforward computations:
\begin{lem}
There exists a unique derivation on $F$ such
that the set of bilinear equations in Theorem \ref{thm:bilinear} holds. 
\end{lem}

Consider the extended field
$\widehat{F}:=F(\tau_0,\tau_1,\tau_2,\sigma_1,\sigma_2)$ 
with a relation
$$
 \tau_1'\tau_2-\tau_2\tau_1'=\sigma_1\sigma_2.
$$
We can naturally extend the derivation by $\tau_j'=g_j'\tau_j$,
$\sigma_k'=h_k'\sigma_k\;(j=0,1,2,k=1,2)$.
Then we have the previous presentation (\ref{Fhat}).
Now the next lemma is a direct consequence of Theorem \ref{thm:bilinear}.
\begin{lem}
We have a natural embedding of the differential fields
$$
 K \subset \widehat{F}.
$$
\end{lem}
\bigskip

Our next task is to extend the affine Weyl group
action on $\widetilde{K}= K(\tau_0,\tau_1,\tau_2,\sigma_1,\sigma_2)$
(Theorem \ref{thm:tauaut}) to $\widehat{F}$.
The following two lemmas can be easily verified.
\begin{lem} \label{lem:Waction}
By the following 
formulae, we can introduce an action of the affine Weyl group $W$ 
on $\widehat{F}$ as a group of automorphisms:
\begin{align*}
 \frac{s_0(\tau_0')}{s_0(\tau_0)}
&=\frac{\tau_0'}{\tau_0}-\frac{\alpha_0}{f_0}, 
\\
 \frac{s_1(\tau_1')}{s_1(\tau_1)}
&=\frac{\tau_1'}{\tau_1}-\frac{\alpha_1}{f_1}\frac{\sigma_2}{\tau_2},
\quad
 \frac{s_1(\sigma_1')}{s_1(\tau_1)}
=\frac{\sigma_1'}{\tau_1}-\frac{\alpha_1}{f_1}\frac{\tau_0'}{\tau_0},
\\
 \frac{s_2(\tau_2')}{s_2(\tau_2)}
&=\frac{\tau_2'}{\tau_2}+\frac{\alpha_2}{f_2}\frac{\sigma_1}{\tau_1},
\quad
 \frac{s_2(\sigma_2')}{s_2(\tau_2)}
=\frac{\sigma_2'}{\tau_2}-\frac{\alpha_1}{f_1}\frac{\tau_0'}{\tau_0},
\end{align*}
and $s_i(\tau_j')=\tau_j'\;(i\ne j),\;
 s_i(\sigma_k')=\sigma_k'\;(i\ne k).$ 
Moreover this action is an extension of the action 
of $W$ on $\widetilde{K}$.
\end{lem}

\begin{lem} \label{1stdertau} 
For $i,j=0,1,2$ and $k=1,2$ we have
$$
s_i(\tau_j')=s_i(\tau_j)',\quad
s_i(\sigma_k')=s_i(\sigma_k)'.
$$
\end{lem}

\textbf{Remark.} Although we have introduced the Weyl group action on the
$\tau$-functions
in an ad hoc manner, these formulae can be derived 
systematically by using the gauge matrices $G_i$ (\ref{Gi}), if 
we identify the $\tau$-functions with the 
components of a {\it dressing matrix}.
We will give an explanation of this point in
a separate article.

\bigskip

The goal of this section is the
following fact:
\begin{thm}{\label{last}}
The derivation of $\widehat{F}$ commutes
with the action of $W$ on $\widehat{F}.$
\end{thm}
A straightforward verification of this fact may require quite a bit of 
calculations, because
the second derivatives of $\tau$-functions
are determined implicitly by the bilinear equations. 
To avoid the complexity, we 
make use of the fact $\widehat{F}={\widetilde{K}}(k)$,
which is easily seen from (\ref{defoftau}), (\ref{u2}), and (\ref{u1}), 
where we set
$$
 k = 2\left(\frac{\tau_0'}{\tau_0}
+\frac{\tau_1'}{\tau_1}
+\frac{\tau_2'}{\tau_2}
\right)
+\frac{\sigma_1'}{\sigma_1}
+\frac{\sigma_2'}{\sigma_2}.
$$
As for the first derivatives of $\tau$-functions, we have already
lemma \ref{1stdertau}. 
Therefore, in order to prove Theorem \ref{last}, it suffices to show the
next lemma.
\begin{lem}
\beqn
  s_i(k')=s_i(k)'\quad(i=0,1,2).
\label{lastlem}
\eeqn
\end{lem}
\begin{proof}
By Lemma \ref{lem:Waction}, we have
\begin{align}
  s_0(k) - k
&= 2\left(\frac{s_0(\tau_0')}{s_0(\tau_0)} 
 -\frac{\tau_0'}{\tau_0} \right) 
 = -2\frac{\alpha_0}{f_0},
\nonumber \\
  s_1(k)-k
&= 2\left(\frac{s_1(\tau_1')}{s_1(\tau_1)}-\frac{\tau_1'}{\tau_1}\right)
  + \left(\frac{s_1(\sigma_1')}{s_1(\sigma_1)}
  - \frac{\sigma_1'}{\sigma_1}\right),
\label{s1k} \\
  s_2(k)-k
&= 2 \left(\frac{s_2(\tau_2')}{s_2(\tau_2)}- \frac{\tau_2'}{\tau_2}\right)
+ \left(\frac{s_2(\sigma_2')}{s_2(\sigma_2)}
- \frac{\sigma_1'}{\sigma_1}\right).
\label{s2k}
\end{align}
We can rewrite the right hand sides of \eqref{s1k} and \eqref{s2k}
into 
\begin{align*}
  s_1(k)-k
&= -2\frac{\alpha_1}{f_1}r
   -\frac{\alpha_1(2xq - f_2)}{3q(f_1q + \alpha_1)},
\\
  s_2(k)-k
&= -2\frac{\alpha_2}{f_2}q
   -\frac{\alpha_2(2xr - f_1)}{3r(f_2r - \alpha_2)}
\end{align*}
by using \eqref{defoftau}, \eqref{s1tau} and \eqref{s2tau}.
On the other hand, the normalization condition (\ref{NorTau}) reads
\begin{align*}
   k'=&-u_0^2 - u_2^2 
       - \left( u_0 - \frac{f_1}{3r} + \frac{2x}{3} \right)^2
       - \left( u_2 - \frac{f_2}{3q} + \frac{2x}{3} \right)^2  
\\
      & \quad + \frac{2x}{9}\left( {4x} - \frac{f_1}{r} - \frac{f_2}{q} \right)
       +\frac{\alpha_1-\alpha_2}{9}.
\end{align*}
Then we can verify (\ref{lastlem}) by
applying (\ref{compati_G}) to $s_i(k')$ and the 
ODE (\ref{compati}) to $s_i(k)'$.
\qed
\end{proof}

\section{Discussion}

We have derived a two-parameter family of the fifth Painlev\'e equation
as a similarity reduction of the modified Yajima-Oikawa hierarchy,
which is related to a non-standard Heisenberg subalgebra
of $A_2^{(1)}.$
The system admits a group of B\"acklund transformations of type
$W(A_2^{(1)}).$
By a suitable modification of
our construction, it may be possible to recover a {\it missing} parameter
and get the fifth Painlev\'e with the full symmetry
of type $W(A_3^{(1)}).$
Combinatorial and/or representation theoretical structure
of the hierarchy is also deserves to be investigated.
A combinatorial aspect of representation associated 
with the Yajima-Oikawa 
hierarchy is studied by S. Leidwanger in \cite{Leid}.
It seems that the work is closely related
some family of polynomial solutions of  
the fifth Painlev\'e equation.
We hope that we discuss these issues in future publications.

\section*{Acknowledgments}
The authors are grateful to 
Masatoshi Noumi, Yasuhiko Yamada, Kanehisa Takasaki,
Koji Hasegawa, Gen Kuroki and Ralph Willox
for fruitful discussions and kind interest.

%

\end{document}